
%
%
\def\afterperiod{\hskip 4.44444pt plus 4.99997pt minus 0.37036pt}
\def\aftercomma{\hskip 3.33333pt plus 2.08331pt minus 0.88889pt}
\baselineskip=16pt
\bigskip
\centerline{\bf A FAST ALGORITHM FOR}
\centerline{\bf RANDOM SEQUENTIAL ADSORPTION OF DISCS}
\bigskip
\centerline{JIAN--SHENG WANG\footnote*{E-mail address:
cscwjs@leonis.nus.sg}}
\smallskip
\centerline{\sl Computational Science, Blk S16,
National University of Singapore,}
\centerline{\sl Lower Kent Ridge Road, Singapore 0511}
\bigskip

\centerline{ABSTRACT}

{\narrower
An efficient simulation algorithm for random sequential
adsorption of discs is implemented.  By dividing the surface
into small squares, depositions are attempted only on squares
that are actually possible.  A crucial part is a method to
identify the available squares.  A precise value of the jamming
coverage is obtained: $\theta(\infty) \approx 0.547\,069$.
The asymptotic law $\theta(t) \approx
\theta(\infty) - c\,t^{-1/2}$ is verified to a high accuracy.
The pair correlation function at the jamming state is
analyzed.\par}

\bigskip
{\noindent\bf 1. Introduction}

Random sequential adsorption (RSA) is an irreversible
process which models depositions of large molecules (e.g.
proteins) on surfaces.$^1$\afterperiod In RSA large
extended objects are deposited on a flat surface one by one,
with locations chosen at random.  The deposited objects are
fixed on the surface and exclude
certain region for further deposition, i.e., the objects are not
allowed to overlap.  A simple model is the deposition of hard
discs on a two-dimensional
continuum.$^{2-9}$\afterperiod Depositions of
squares,$^{8,10-12}$\aftercomma elongated objects like ellipses
and needles,$^{13-17}$\aftercomma mixtures of different
shapes,$^{18,19}$\aftercomma as well as lattice
models$^{8,12,20-23}$ were also studied by many authors.
Experiments on RSA lagged behind theoretical studies, but
beginning to catch up.$^{24,25}$

RSA exhibits several interesting features not present in
equilibrium systems.  First of all, there exists a jamming
state, which has the highest possible particle density smaller
than the close-packed density of the corresponding equilibrium
system.  At the jamming state, the distances between objects are
such that no other object can be added to the surface.  The
jamming density is approached via a power law on continuum and
an exponential law on discrete lattices.  At the jamming state, the
pair correlation function diverges logarithmically when
approaching the distance of contact.

The coverage (or density) at large time $t$ takes the form
$$ \theta(t) \approx \theta(\infty) - {c \over t^{1/d_f}}, \eqno(1)$$
where $c>0$ is some constant, the exponent $d_f$ depends on the
geometry of the objects.  The most interesting quantities are
the value $\theta(\infty)$ and the scaling exponent $d_f$.
Analytic arguments$^{26,27}$ and numerical work$^{4,6}$ support
the result that $d_f = d$, for hyperspheres in $d$ dimensions.
In two dimensions the jamming limit is approaches by $ c /
\sqrt{t}$ for discs.  This result is referred as Feder's law in
the literature.  For anisotropic objects like ellipses and
unoriented squares,$^{10,13,16}$\aftercomma the jamming coverage
is approached even slower, $d_f = 3$.  It is argued that $d_f$
is the number of degrees of freedom of the object in
general.$^{10}$\afterperiod On the other hand, for the
deposition of squares with fixed orientation (oriented squares),
the convergence law is$^{11,12,27}$ $\ln t/t$.

In this paper, we present an efficient method of the RSA simulation
and report the results of extensive runs on a cluster of
workstations.

\medskip
{\noindent\bf 2. Standard and Event-Driven Simulation of RSA}

The standard RSA model of hard discs$^{2-6}$ is defined as
follows: We begin with an empty, flat surface of $L \times L$ at
time $t=0$. A point $(x,y)$ is drawn from a uniform probability
distribution over the entire surface area.  A deposition of a
disc of diameter $\sigma = 1$ with center at $(x,y)$ is
attempted.  If this disc does not overlap with previously
deposited discs, i.e., if the distances between the current
point $(x,y)$ and the centers of all other discs on the surface
are greater than $\sigma$, the deposition is accepted, otherwise
it is rejected.  Each attempt increases the time by $1/L^2$.
The deposition trial is repeated until no more deposition is
possible, at which point, the jamming state is reached.  The
coverage $\theta(t)$ is the fraction of the area covered by the
discs at time $t$.

The standard computer simulation of RSA follows closely the
description of the model.  The checking for overlap can be done
quickly with the help of grids.$^6$\afterperiod The total
$L\times L$ surface area is partitioned into regions of squares
of size $1 \times 1$.  A disc at $(x,y)$ belongs to the unit
square at $(\lfloor x \rfloor,\, \lfloor y \rfloor)$, where the
notation $\lfloor x \rfloor$ represents the floor or integer
part of $x$.  An array of linked lists of disc coordinates is
used to represent the deposited discs.  With this data
structure, it is suffice just to check the discs located at the
center and eight surround squares for overlaps, see Fig.~1.

The above simple method has a severe shortcoming that the study
of late-stage process is very time-consuming.  When the jamming
configuration is about to be reached, most of the area is
blocked.  Only tiny disconnected regions are available for
further deposition.  Since we choose the sites at random with a
uniform probability distribution, the available sites are harder
to find.  To overcome this difficulty, we incorporated two
important ideas: (1) divide the surface into small squares and
make deposition attempts only on squares that are not completely
blocked; (2) systematically reduce the sizes of the small
squares after some number of attempts and re-evaluate the
availability of the squares.  The first strategy is similar to
that for the simulation of Ising model at low
temperatures,$^{28}$ as well as methods of RSA.$^{11,
12}$\afterperiod The second strategy appears new and very
effective.

In this more elaborate method, we make attempts only on squares
which are potentially possible for depositions.  Thus the core
part is an algorithm which identifies correctly whether a square
is available for deposition.  Since the diameter of the discs is
$\sigma$, each disc excludes an area of circle of radius
$\sigma$ for further deposition (Fig.~1).  If some area is
completely covered by a union of the exclusion zones of discs, that
area is not available for deposition.  We begin by classifying
all the squares of $a\times a$ with $a=1$ as available or
unavailable squares.  The available squares are put on a list.
Deposition trials are taken only on the squares in the list, i.e., a
square in the list is chosen at random, deposition is attempted
with a uniform probability over the square. After certain number
of trials (we used $5 \times 10^5$), we rebuild the list, now
with squares of size $(a/2) \times (a/2)$, checking only those
squares on the old list.  Each old square is subdivided into 4
smaller squares.  This shrinking of the basic squares helps to
locate even extremely small available regions.  They will be hit
with very small probability if the area is always $1 \times 1$.
Because of integer overflow,
the size of the squares is not allowed to go arbitrarily small,
but the shrinking is stopped at $ a = 2^{-15}$.  Thus the
smallest square has an area of $2^{-30} \approx 10^{-9}$.

For the standard RSA algorithm, each trial increases the time
$t$ by $1/A$, where $A = L^2$ is the area of the total surface.
In the even-driven algorithm, where only the potentially
successful depositions are tried, the clock must tick faster in
order to compensate for not making attempt in the completely
unsuccessful area.  The time increment for each trial on the
available squares is a random variable,$^{11}$
$$ \Delta t = {1\over A} \left( \left\lfloor { \ln \xi \over
\ln(1 - A_s/A)} \right\rfloor + 1 \right), \eqno(2)$$
where $A_s$ is the total area on which we try our deposition, it is
equal to $a^2$ times the number of available squares; and $\xi$
is a uniformly distributed random number between 0 and 1.  This
interpretation of time interval makes the two algorithms
equivalent.

The above result can be understood in the following way: the
total area is divided into $A_s$ and $A-A_s$.  If we try
to deposit the discs uniformly on the whole area $A$, the
probability that a trial landed on $A_s$ is $p = A_s/A$.  If the
first attempt fails to hit $A_s$, a second attempt landed on
$A_s$ has the probability $p( 1- p)$.  In general, the
probability that the first $i-1$ trials landed in the
unavailable area $A-A_s$ and the $i$-th trial landed on $A_s$ is
$$P_i = p\, ( 1 - p )^{i-1}, \quad i = 1,\,2,\,3, \cdots.  \eqno(3)$$
In the event-driven simulation, from first trial in the area
$A_s$ to a second trial in the same area, the time has elapsed
by $\Delta t = i/A$, since $i-1$ unsuccessful attempts had been
made in the area $A - A_s$, where $i$ is distributed according
to Eq.~3.  This exponential distribution is easily generated by
the logarithmic function, given Eq.~2.

\medskip
{\noindent\bf 3. Identification of Fully Covered Squares}

An important piece of code of our RSA simulation program is the
identification of the squares which are fully covered by the
excluded area of discs.  For these squares we are sure that
depositions on them will be unsuccessful, and thus they'll not be
on the list of candidates for trials.

The algorithm that we have devised is as follows.  Written as a
C programming language function, it returns a nonzero value if
the square is fully covered and a value 0 otherwise.  Part (1)
to~(5) is executed in the order given.

\item{(1)} Find relevant discs to the current square.  A disc is
relevant if the square overlaps with the excluded region of the
disc.  Note that a disc has a diameter $\sigma$ and its
excluded region is a concentric circle of radius $\sigma$.

\item{(2)} If all of the four vertices of the square are inside a
single-disc excluded region, then it is already fully covered
(Fig.~2a).  Function returns with value~1.

\item{(3)} For a full coverage, each vertex of the square must
be at least in one of the excluded region of the relevant discs.
If any one of the vertices is not covered at all by any disc,
then the square is not fully covered (Fig.~2b).  Function
returns with value~0.

\item{(4)} At this point, at least two discs are involved if the
function is not returned.  Mapping the edges of the square to a
one-dimensional line between 0 and 4, we find out all the
line segments which are
covered by the excluded region of discs.  If the four edges of
the square are completely covered by the discs, and we have
exactly two discs, then the square is fully covered (Fig.~2c),
function returns with value~2.  If there are segments not
covered by discs, then we know that the square is not fully
covered (Fig.~2d).  Function returns with value~0.

\item{(5)} If the function is not returned, we know that there
are at least three discs.  Even if all the edges are covered,
the square can still contain uncovered area if three or more
discs are involved.  Three or more discs can form holes.  To
make the last check, the coordinates of the intersection points
of circles (with radius $\sigma$) are calculated.  In order for
the interior of the square to be covered, all the intersection
points of any pair of discs which lie inside or on the square
edges must be covered by a third disc.  Return with the number
of relevant discs if the above condition is satisfied (Fig.~2e);
return value~0 if not (Fig.~2f).

\noindent The order of various checks is arranged in such a way
so that the most frequent situations are checked first. Even
though the worse case computational complicity goes as $n^2$,
where $n$ is the number of relevant discs, a definite conclusion
can be made typically much earlier.

\medskip
{\noindent\bf 4. Simulation Results}

We used mainly a system of linear size $L = 1000$.  This appears to be
the largest system considered in RSA simulation.  With this size
various data occupied 24 megabytes.  Periodic boundary condition
is used.  This is crucial to eliminate an obvious edge effect of
order $1/L$. It is known that finite-size effect in RSA is
extremely small$^{11}$ due to rapid damping of two-point
correlations.  The actual form of the finite-size effect is not
known.  In any case, a system of $1000 \times 1000$ is more than
adequate and finite-size effect can be neglected.

For time $t \leq 8$ we used the standard method.  This method is
simpler, thus faster in the initial deposition process.  The use
of the event-driven algorithm at this stage would require much
more memory.  The event-driven method is used for $ t > 8$.  In
the beginning, the size of small squares (for classification
purpose) is $1 \times 1$, which is the same as the underline
grids for book-keeping of adjacency of discs.  The squares are
subdivided into four squares after every $0.5 L^2$ trials.  This
helps to keep the number of available squares less than $ 0.3
L^2$, even though the area of the squares is smaller.

A simulation terminates if there is no more available square, or
if $t > 2^{49}$ (a complete run).  On a total
area of $1000 \times 1000$, the jamming state is reached
typically at a time $ t \approx 2^{42}$ with large fluctuation.
In general, the jamming state is obtained on the time scale of
$t \approx L^4$.  In each complete process of deposition of
discs on a finite lattice, the coverage varies discretely.  The
last disc deposited on the surface causes a variation in the
coverage by an amount of order $1/L^2$.  Comparing this
variation with the result of an infinitely large system or with
the result of average over many runs, which is of order
$t^{-1/2}$, we conclude that the major deposition process is
finished when these two numbers are of the same order of
magnitude, given $t \approx L^4$.

Each complete run took 22 minutes on an SGI Indigo workstation.
The standard simulation method ran at the speed of about one
step ($\Delta t = 1$, $L^2$ attempts) per minute, independent of
the time $t$.  If we were determined to use this method for all
$t$, a complete run would have taken $2^{42}\,$minutes ($\approx
8\,$ million years) to finish.

Floating-point calculations are carried out in double precision.
We used two random number generators in the program, a 64-bit
linear congruential generator$^{29}$ and a 48-bit one
implemented on many C programming language environment on
workstations.  Both of them take the form
$$ x_{n+1} = a\, x_n + c \bmod m, \eqno(4)$$
with $a = 6\,364\,136\,223\,846\,793\,005$, $c=0$, $m=2^{64}$
for the 64-bit random number and $a = 25\,214\,903\,917$,
$c=11$, $m=2^{48}$ for the 48-bit random number.

The coverage data are taken at $t = 2^k$, with $k = 0$, 1, 2,
$\ldots$, 49.  The pair correlation function is calculated from
the last configuration.  The final results are averages over
$43\,427$ runs.

To check the $1/\sqrt{t}$ convergence law of the coverage, we
plot in Fig.~3 $\log_2 \bigl[\theta(2t) - \theta(t) \bigr]$ vs.\
$\log_2 t$.  Assuming a power-law dependence of the form in
Eq.~1, we have
$$ \log_2 \bigl[ \theta(2t) - \theta(t) \bigr] =
c_0 - {1\over d_f } \log_2 t, \eqno(5)$$
where $c_0 = \log_2 \bigl[ c\, ( 1 - 2^{-1/d_f} ) \bigr]$.  By a
linear least-squares fit, we find that $d_f = 2.0008 \pm 0.0016$
and $c = 0.300 \pm 0.002$, confirming the theoretical
expectation ($d_f=2$) to high accuracy.

To get the jamming coverage, we can simply take the value at
$t=2^{49}$.  The actual difference between $t=2^{49}$ and
$t\to\infty$ is in fact much smaller than the statistical
errors.  The estimate for the jamming coverage is thus
$$ \theta(\infty) = 0.547\,069\,0 \pm 0.000\,000\,7. \eqno(6) $$
This result agrees with previous work,$^{2-4,6,19}$\aftercomma
$\theta(\infty) \approx 0.547$, but is two to three orders of
magnitude more accurate.

The pair correlation function of the final configuration (at $t=
2^{49}$) is presented in Fig.~4.  The pair correlation function
$g(r)$ is defined as the average number of discs per unit area
located at a distance $r$ away, given that there is a disc at
the origin, normalized by the jamming density so that $g(\infty)
= 1$.  A striking feature is the divergence at $ r \to 1$.
Generalizing an exact result in one dimension, Pomeau$^{26}$ and
Swendsen$^{27}$ showed that
$$ g(r) \approx a \ln (r - 1)  + b. \eqno(7) $$
The logarithmic divergence is nicely confirmed from our
numerical data (Fig.~4, insert).  A least-squares fit with data
$\ln(r-1) < -6$ gives $a =
-1.535 \pm 0.002$ and $b = -2.13 \pm 0.02$.

\medskip
{\noindent\bf 5. Conclusion}

We proposed a very efficient simulation algorithm for the random
sequential adsorption of discs.  With this new method and
extensive runs, accurate jamming coverage is obtained.  The
Feder's law for the large time asymptotic coverage and logarithmic
divergence of the pair correlation are confirmed
to a high accuracy.  The method can be easily generalized to RSA
of other geometric shapes with fixed orientation.  It appears
complicated to apply to problems where the objects have a
rotational degree of freedom, e.g., deposition of ellipses.

\vfill\eject
{\noindent\bf References}
\medskip

{\frenchspacing

\item{1.}  For recent reviews see: M. C. Bartelt and V. Privman,
Int. J. Mod. Phys. B {\bf 5}, 2883 (1991); J. W. Evans, Rev. Mod. Phys.
{\bf 65}, 1281 (1993).

\item{2.} L. Finegold and J. T. Donnell, Nature, {\bf 278}, 443 (1979).

\item{3.} M. Tanemura, Ann. Inst. Stat. Math. {\bf 31}, 351 (1979).

\item{4.} J. Feder, J. Thoer. Biol. {\bf 87}, 237 (1980).

\item{5.} J. Feder and I. Giaever, J. Colloid Interface Sci. {\bf 78},
144 (1980).

\item{6.} E. L. Hinrichsen, J. Feder, and T. J\o ssang, J. Stat. Phys.
{\bf 44}, 793 (1986).

\item{7.} P. Schaaf and J. Talbot, Phys. Rev. Lett. {\bf 62},
175 (1989).

\item{8.} R. Dickman, J.-S. Wang, and I. Jensen, J. Chem. Phys.
{\bf 94}, 8252 (1991).

\item{9.} J. A. Given, Phys. Rev. A {\bf 45}, 816 (1992).

\item{10.} P. Viot and G. Tarjus, Europhys. Lett. {\bf 13}, 295 (1990).

\item{11.} B. J. Brosilow, R. M. Ziff, and R. D. Vigil, Phys. Rev. A
{\bf 43}, 631 (1991).

\item{12.} V. Privman, J.-S. Wang, and P. Nielaba, Phys. Rev. B {\bf 43},
3366 (1991).

\item{13.} R. D. Vigil and R. M. Ziff, J. Chem. Phys. {\bf 91}, 2599 (1989).

\item{14.} J. D. Sherwood, J. Phys. A: Math. Gen. {\bf 23}, 2827 (1990).

\item{15.} R. M. Ziff and R. D. Vigil, J. Phys. A. Math. Gen.
{\bf 23}, 5103 (1990).

\item{16.} P. Viot, G. Tarjus, S. M. Ricci, and J. Talbot, J. Chem. Phys.
{\bf 97}, 5212 (1992).

\item{17.} S. M. Ricci, J. Talbot, G. Tarjus, and P. Viot, J. Chem. Phys.
{\bf 97}, 5219 (1992).

\item{18.} G. Tarjus and J. Talbot,  J. Phys. A: Math. Gen.
{\bf 24}, L913 (1991).

\item{19.} P. Meakin and R. Jullien, Phys. Rev. A {\bf 46}, 2029 (1992).

\item{20.} R. S. Nord and J. W. Evans, J. Chem. Phys. {\bf 82}, 2795 (1985).

\item{21.} P. Meakin, J. L. Cardy, E. Loh, and D. J. Scalapino,
J. Chem. Phys. {\bf 86}, 2380 (1987).

\item{22.} P. Schaaf, J. Talbot, H. M. Rabeony, and H. Reiss,
J. Chem. Phys. {\bf 92}, 4826 (1988).

\item{23.} Y. Fan and J. K. Percus, Phys. Rev. Lett. {\bf 67}, 1677 (1991).

\item{24.} G. Y. Onoda and E. G. Liniger, Phys. Rev. A {\bf 33}, 715 (1986).

\item{25.} J. J. Ramsden, Phys. Rev. Lett. {\bf 71}, 295 (1993).

\item{26.} Y. Pomeau, J. Phys. A {\bf 13}, L193 (1980).

\item{27.} R. H. Swendsen, Phys. Rev. A {\bf 24}, 504 (1981).

\item{28.} A. B. Bortz, M. H. Kalos, and J. L. Lebowitz, J. Comp. Phys.
{\bf 17}, 10 (1975).

\item{29.} D. E. Knuth, in {\sl the Art of Computer Programming},
Vol. 2, 2nd ed. (Addison-Wesley, 1981).

}  

\vfill\eject
{\noindent\bf Figure Captions}
\medskip

\noindent\hang{\bf Fig.~1.}\ \ RSA configuration with five discs
deposited on surface.  Each disc has an exclusion zone (black
and grey region).  The grids are used to locate the neighboring
discs.

\medskip
\noindent\hang{\bf Fig.~2.}\ \ Various geometric relations
between the square and excluded regions of discs.  In
situation (a), (c), or~(e) the square is fully covered by one, two, or
more than two discs; (b), (d), or~(f) fails to cover the square.

\medskip
\noindent\hang{\bf Fig.~3.}\ \ The difference in coverage,
$\theta(2t) - \theta(t)$, vs.~time $t$, plotted in double
logarithmic (to the base 2) scale.  The data points are simply
connected by straight lines.

\medskip
\noindent\hang{\bf Fig.~4.}\ \ Pair correlation function $g(r)$ at the
jamming state.  Insert is a portion near $ r \to 1$.

\bye